\pdfoutput=1
\documentclass[pdftex,twocolumn,epjc3]{svjour3}          % twocolumn

\RequirePackage[T1]{fontenc}

\smartqed  % flush right qed marks, e.g. at end of proof

\RequirePackage{graphicx}
\RequirePackage{mathptmx}      % use Times fonts if available on your TeX system
\RequirePackage{flushend}
\RequirePackage[numbers,sort&compress]{natbib}
\RequirePackage[colorlinks,citecolor=blue,urlcolor=blue,linkcolor=blue]{hyperref}

\journalname{Eur. Phys. J. C}
\usepackage{dcolumn}
\usepackage{color}
\usepackage{booktabs}
\usepackage{xspace}
\usepackage{graphicx}
\usepackage{amsmath}
\usepackage{amssymb}
\usepackage[figuresright]{rotating}
\usepackage{subfigure}
\usepackage{lineno}

\graphicspath{
 {IMAGES/}
}
\newcommand{\TeVe}         {Te\kern-.1emV\xspace}
\newcommand{\GeVe}         {Ge\kern-.1emV\xspace}

\begin{document}
\title{
Extraction of the specific shear viscosity of quark-gluon plasma from two-particle transverse momentum correlations
}
\author{Victor Gonzalez \thanksref{e1,addr1} \and
Sumit Basu \thanksref{e2,addr2} \and
Ana Marin \thanksref{e3, addr3} \and
Jinjin Pan \thanksref{addr4} \and
Pedro Ladron de Guevara \thanksref{addr5,addr6} \and
Claude A. Pruneau \thanksref{e4, addr1}
}

\thankstext{e1}{e-mail: {\em victor.gonzalez@cern.ch}}
\thankstext{e2}{e-mail: {\em sumit.basu@cern.ch}}
\thankstext{e3}{e-mail: {\em a.marin@gsi.de}}
\thankstext{e4}{e-mail: {\em claude.pruneau@wayne.edu}}

\institute{Department of Physics and Astronomy, Wayne State University, Detroit, Michigan 48201, USA\label{addr1} \and 
Lund University, Department of Physics, Division of Particle Physics, Box 118, SE-221 00, Lund, Sweden\label{addr2} \and 
GSI Helmholtzzentrum f\"ur Schwerionenforschung, Research Division and ExtreMe Matter Institute EMMI, Darmstadt, Germany\label{addr3} \and 
Cyclotron Institute, Texas A\&M University, College Station, Texas 77843, USA\label{addr4} \and 
Universidad Complutense de Madrid, Spain\label{addr5} \and 
Instituto de F\'{\i}sica, Universidad Nacional Aut\'{o}noma de M\'{e}xico, CP 04510, CDMX, Mexico\label{addr6}}

\date{Received: date / Revised version: date}
\maketitle

\begin{abstract}
The specific shear viscosity, $\eta/s$, of the quark-gluon plasma formed in ultrarelativistic heavy-ion collisions at RHIC and LHC is estimated  based on the progressive longitudinal broadening of transverse momentum two-particle correlators, $G_2$, reported  as a function of collision centrality by the STAR and ALICE experiments. Estimates are computed as a function of collision centrality using the Gavin ansatz which relates the $G_2$ longitudinal 
broadening to the specific shear viscosity. Freeze out times required for the use of the ansatz are computed using  a linear fit of  freeze out times  reported as  a function of the cubic root of the charged particle pseudorapidity density (${\rm d}N_{\rm ch}$/d$\eta)^{1/3}$. Estimates of $\eta/s$
based on ALICE data exhibit little to no dependence on collision centrality at LHC energy, while estimates obtained from STAR data hint that $\eta/s$ might be a function of collision centrality at top RHIC energy. 
\end{abstract}

\section{Introduction}
A key focus of the ultrarelativistic heavy-ion collision programs  conducted at the Large Hadron Collider (LHC)
and the Relativistic Heavy-Ion Collider (RHIC) involves precision measurements  of the  properties of the
quark-gluon plasma (QGP) formed in high-energy nucleus-nucleus collisions. Of particular interest are the magnitude and
temperature dependence of the specific shear viscosity of the QGP, expressed as the ratio $\eta/s$ of
the shear viscosity $\eta$ to the entropy density $s$ of the matter produced in the collisions.  Shear viscosity  
characterizes the ability of a medium to transport momentum and carry deformations.  Transverse particle
anisotropy patterns, quantified in terms of  anisotropic flow coefficients, measured  in mid central heavy-ion
collisions at RHIC and LHC are rather large and were, from the onset, relatively well reproduced by  viscosity free
hydrodynamical  calculations thereby suggesting the QGP might be a perfect fluid, i.e., a fluid with vanishing or
negligible shear 
viscosity~\cite{Shuryak:2003xe,Romatschke:2007mq,Luzum:2008cw,Bozek:2009dw,Song:2010mg,Shen:2011eg}. The possibility that the high temperature,
high density systems formed in the midst of heavy-ion collisions might be a perfect fluid thus generated `quite' a bit 
of excitement~\cite{bnl:20050418}. Considerable experimental and theoretical  efforts were consequently 
expanded to determine the specific shear viscosity of the matter produced at RHIC and more 
recently at the
LHC~\cite{Gale:2013da,Heinz:2013th} based on measurements of anisotropic flow in the collision transverse plane. Although theoretical  efforts have been quite successful in reducing the range of $\eta/s$ compatible with state of the art measurements of flow anisotropies, there  still remains a certain degree of ambiguities owing to several technical difficulties. One of these technical difficulties involves the lack of knowledge on the initial conditions of the systems produced in A--A collisions at RHIC and LHC \cite{Song:2010mg,Qiu:2011hf,Song:2012tv} . For instance, CGC inspired initial conditions yield larger initial spatial anisotropy than MC Glauber type initial conditions and thus require a somewhat larger level of shear viscosity to match the observed flow coefficients when used as input to viscous hydrodynamics simulations. While  efforts to reduce the initial conditions ambiguity based on measurements of symmetric cumulants \cite{ALICE:2016kpq,Niemi:2015qia}, in particular, have had some success, it remains of interest to identify techniques that might enable measurements of specific shear viscosity that are less susceptible to uncertainties associated with initial conditions.
Such a technique exists. Proposed by Gavin \textit{et al.} already more than a decade ago~\cite{Gavin:2006xd}, it involves  measurements of the longitudinal broadening of a  transverse momentum two particle correlator, now dubbed $G_2$,  with increasing collision centrality. The correlator $G_2$, defined in Ref.~\cite{Gavin:2006xd,Sharma:2008qr}, is designed to be proportional to the covariance of momentum currents and is as such sensitive to dissipative viscous forces at play during the transverse and longitudinal expansion of the matter formed in A--A collisions. Gavin \textit{et al.} showed these forces lead to a longitudinal broadening of $G_2$ measured as a function of the pseudorapidity difference of measured charged particles. As the matter expands, neighboring fluid cells drag one another. Fast fluid cells tend to slow down whereas slow fluid cells accelerate.  This has the effect of dampening the expansion and produces a progressive broadening of the $G_2$ correlator with time. The longer the system lives, the longer viscous effects play a role, and the broader the $G_2$ correlator  becomes. Gavin \textit{et al.} showed the broadening, characterized in terms of the difference of the variance of the correlator observed in most central and most peripheral collisions, should be proportional to $\eta/s$ and given by the following formula herein called the Gavin ansatz

\begin{equation}
    \sigma_{\rm c}^{2} - \sigma_{0}^{2} = \frac{4}{T_{\rm c}}
        \,\frac{\eta}{s}\,\left(\frac{1}{\tau_{0}} - \frac{1}{\tau_{\rm c,f}}\right),
  \label{eq:sigmacentral}
\end{equation}
where $\sigma_{\rm c}$ is the longitudinal width of the correlator measured in  most central collisions 
whereas $\sigma_{0}$ is the longitudinal width of the correlator at formation time $\tau_{0}$. $T_{\rm c}$ and $\tau_{\rm c,f}$ are respectively the critical temperature 
and the freeze-out time in most central collisions. 

We first briefly review, in sec. 2, prior efforts to determine $\eta/s$ based on the longitudinal broadening of the $G_2$ correlator in A--A collisions. The method and results of this work are presented in sec. 3 and discussed in sec. 4. Our conclusions are presented in sec. 5.

\section{\texorpdfstring{Prior estimates of $\eta/s$ based on transverse momentum correlations}{Prior estimates of eta/s based on transverse momentum correlations}}

A first estimate of the QGP viscosity based on the Gavin ansatz was reported several years ago by  the STAR collaboration using   a measurement of the charge independent correlator $G_2^{\rm CI}$  in Au--Au collisions at $\sqrt{s_{\rm NN}} = 0.2\;\text{TeV}$. STAR observed the longitudinal width of the $G_2$ correlator grows considerably from most peripheral to most central Au -- Au collisions. Given the observed  broadening might  arise in part from other dynamical effects, STAR used the Gavin ansatz to estimate  an upper limit and reported  $\eta/s$ to be in  the range  0.06 - 0.21~\cite{Agakishiev:2011fs}. More recently, the ALICE collaboration reported precise measurements of the evolution of the longitudinal and azimuthal widths of  charge independent and charge dependent two-particle transverse momentum correlators,  $G_2^{\rm CI}$ and $G_2^{\rm CD}$, respectively, as a function of the centrality of Pb -- Pb collisions at $\sqrt{s_{\rm NN}} = 2.76$  TeV~\cite{Acharya:2019oxz}. Examining specifically the overall change of the correlator longitudinal  width from most peripheral to most central collisions, the  collaboration concluded that their  observations  favour  small values of $\eta/s$, that is, values  close to the KSS  bound of $1/4\pi$~\cite{Kovtun:2004de}.

The $\eta/s$ estimates reported by the ALICE and STAR collaborations focused on the overall change of the longitudinal width $G_2^{\rm CI}$ from peripheral to central collisions but did not utilize correlator widths observed in mid-central collisions.
This omission resulted in large part from the lack of precise estimates of the system's life time in mid-central collisions. Effectively, STAR and ALICE  did not consider the possibility that the viscosity might evolve with collision centrality and thus did not apply the Gavin ansatz to intermediate ranges of collision centralities.  We note, however, that the viscosity might in fact become a function of the collision centrality if, in particular, the temperature or density of the produced system  or other conditions affecting the viscosity  evolve with  centrality. It is also conceivable that other aspects of the collision dynamics, not related to viscous effects, could impact the broadening of the $G_2$ vs. centrality. It is thus of interest to consider what  the evolution of the $G_2$ correlator observed by STAR and ALICE implies. Two specific questions arise. The first is concerned  purely with the  experimental technique used to estimate $\eta/s$ while the second concerns  a possible evolution of the effective shear viscosity of the system with collision centrality.

Let us first consider the experimental technique on its own merits. Is the technique sound? Are there experimental artifacts that can bias or skew the evaluation of $\eta/s$ based on the Gavin ansatz? Indeed, the ansatz requires estimates of a critical temperature $T_{\rm c}$, as well as initial (formation) and freeze-out times $\tau_0$ and $\tau_{\rm c,f}$, respectively. These quantities are not evaluated in the context of the $G_2$ measurement and thus require external inputs. They may thus be subjected to systematic bias of their own and independent of the STAR and ALICE measurements of the $G_2$ correlator. Additionally, estimation of the broadening of the correlator might perhaps be biased by the finite acceptance or other artifacts of the measurement process. One might wonder, in particular, whether the width observed in most central collisions could be underestimated because of the finite rapidity width of the acceptance of the measurements. In this context, it becomes of interest to study what progressive changes of the width might imply about the strength of the specific shear viscosity, and whether, in particular, the evolution of the widths with centrality is self-consistent, that is, whether changes of the width from one fractional cross section to the next are consistent with the overall change from most peripheral to most central collisions.  

The second set of concerns is of greater interest, from a physical standpoint, but perhaps more difficult to elucidate. Are viscous effects strictly proportional to the system lifetime? Can the correlator be affected by other physical effects, such as, possibly, the radial and anisotropic expansion of the collision system? Is the characteristic temperature used in the ansatz truly a constant independent of the collision centrality? And perhaps, most interestingly, could the effective shear viscosity extracted from the measurement be a function of collision centrality? Theoretical considerations suggest $\eta/s$ is likely a function of the QGP  temperature~\cite{Denicol:2015nhu,Okamoto:2017rup,Niemi:2011ix,Niemi:2015qia,Molnar:2014zha,Haas_2014,Christiansen_2015,Bernhard:2019bmu}. Is it then possible that collisions at different impact parameter yield systems at different temperatures with slightly different time evolution  of the shear viscosity, thereby resulting in effective or time-averaged shear viscosity that might depend on the collision centrality? Conceivably,  answers to these questions may require more and better data than those available, but it is nonetheless of interest to consider what the available data can say about a possible evolution of $\eta/s$ with collision centrality and system temperature. It is thus the primary objective of this work to explore how $\eta/s$ values obtained with the Gavin ansatz evolve with collision centrality.

\section{\texorpdfstring{Evolution of $\eta/s$ with system size}{Evolution of eta/s with system size}}
\begin{figure}[hbt]
    \centering
    \includegraphics[width=\columnwidth,keepaspectratio=true,clip=true,trim=20pt 4pt 50pt 30pt]{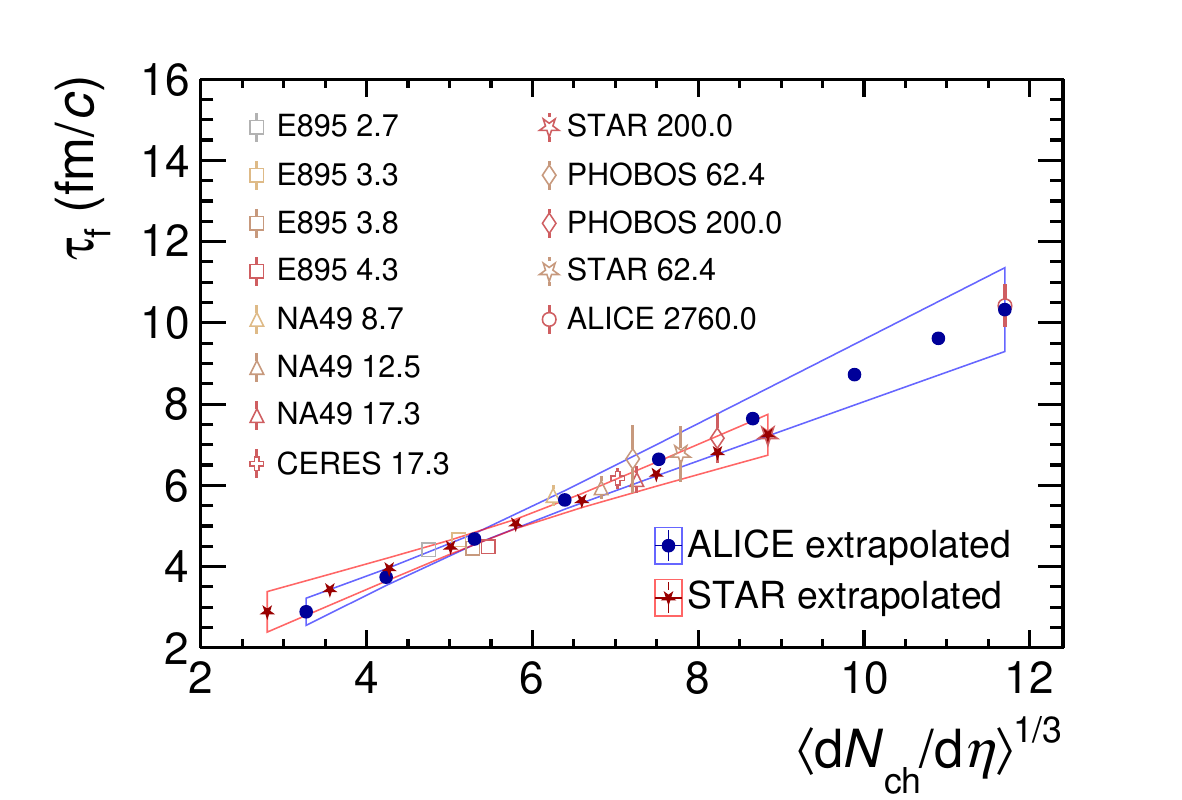}
    \caption{Open symbols: Compilation of decoupling (freeze-out) times, $\tau_{\rm f}$, plotted as a function of the cubic root of the charged particle density, $({\rm d}N_{\rm ch}/{\rm d}\eta)^{1/3}$, observed in various collision systems and for a wide range of beam energies~\cite{Aamodt:2011mr}. Filled symbols: extrapolated values of $\tau_{\rm f}$  corresponding to each of the centrality classes used in our computation of $\eta/s$ based on   Au -- Au collisions at $\sqrt{s_{\rm NN}} =0.2$ TeV and Pb -- Pb collisions at $\sqrt{s_{\rm NN}} =2.76$ TeV, measured by the STAR and ALICE collaborations, respectively. Red/blue solid lines represent systematic uncertainties of the polynomial fits  to the data.}
    \label{fig:freezeout}
\end{figure}

We proceed with the evaluation of  $\eta/s$ as a  function of the cubic root of the pseudorapidity density ${\rm d}N_{\rm ch}$/{\rm d}$\eta$, based  on the $G_2^{\rm CI}$ longitudinal widths already reported by the STAR and ALICE collaborations~\cite{Agakishiev:2011fs,Acharya:2019oxz} in  Au--Au collisions at  $\sqrt{s_{\rm NN}} = 0.2\;\text{TeV}$, and in Pb--Pb at $\sqrt{s_{\rm NN}} = 2.76\;\text{TeV}$, respectively, as a function of the collision centrality using Eq.~\eqref{eq:sigmacentral}. 
However, we also need estimates of the lifetimes $\tau_{\rm f}$ of the system with collision centrality. Estimates of freeze-out times reported as a function of measured charged particle densities, d$N_{\rm ch}$/d$\eta$, in \cite{Aamodt:2011mr}, are used. Values of $\tau_{\rm f}$ are obtained from two-pion Bose-Einstein measurements from AGS to LHC energies~\cite{Aamodt:2011mr}.
Freeze-out times relevant for each of the centrality ranges considered in this work are obtained by fitting a first degree polynomial to estimated values  $\tau_{\rm f}$ according to 
\begin{equation}\label{eq:tau}
    \tau_{\rm f} = A \cdot ({\rm d} N_{\rm ch}/{\rm d}\eta)^{1/3}. 
\end{equation}
Freeze-out times and associated uncertainties are listed in Tab.~\ref{tab:paramForeta} and plotted in  Fig.~\ref{fig:freezeout}.
Statistical and systematic errors reported by the E895, CERES, NA49, PHOBOS, STAR and ALICE collaborations \cite{Aamodt:2011mr,Adams:2004yc,Adamova:2002wi,Lisa:2000hw,Alt:2007uj,Afanasiev:2002mx,Abelev:2009tp,Back:2004ug,Back:2005hs,Back:2002wb,Abelev:2008ab} are used in the least square fit procedure. Note, however, that we could not obtain a  fit with Eq.~(\ref{eq:tau}) that satisfactorily match all available data. We thus proceeded to use linear polynomial fits ($\tau = a_0 + a_1 {\rm d}N_{\rm ch}/{\rm d}\eta$) and  opted to give larger emphasis and weights to $\tau$ estimates obtained at  0.2 and 2.76 TeV given our goal is to  determine  $\eta/s$ based on $G_2$ data acquired at these two energies. Several distinct fits were  carried out to obtain parameterizations of the $\tau_{\rm f}$ dependence on the charged particle density used in our determination of $\eta/s$.  Fit conditions were varied: our primary fit included all data points but we also considered fits based on data in selected energy ranges, and with or without constraining the fits to pass through the origin. All fits considered yield chi-square per degrees of freedom of the order of  $\chi^{2}/{\rm ndf} = 2$. For the STAR energy, 0.2 TeV, the fit that better reproduces the published results does not pass through the origin and yields $A=0.72$ while for the ALICE energy, 2.76 TeV, the best fit yields a straight line that  passes through the origin and a value $A = 0.88$. Fig.~\ref{fig:freezeout} shows the extrapolated decoupling times  $\tau_{\rm f}$  corresponding to the charged particle density used in the measurements of the $G_2$ correlator. Error bands show the systematic uncertainties introduced by the fit procedure.  Additionally, we also carried out  estimates of  $\eta/s$ based on $\tau_{\rm f}$ values obtained from pion interferometry and blast wave fits to particle spectra~\cite{Adams:2004yc,Adams:2003xp,Adam:2015vna,Abelev:2013vea} and found that estimates  $\eta/s$ obtained with these alternative values of $\tau_{\rm f}$ were in agreement, within uncertainties, with the results obtained with $\tau_{\rm f}$ values obtained with the fit procedure described above.

Computation of the Gavin ansatz is accomplished using the canonical values $T_{\rm c} = 160 \pm 5\;\text{MeV}$ and $\tau_{0} = 1.0 \pm 0.5\;\text{fm}/c$  for the
critical temperature and formation time, respectively~\cite{Becattini:2014rea}.
The value of $\sigma_{0}$ is estimated by extrapolating the width of the correlator to $\langle N_{\rm part} \rangle = 2$.
Values of (d$N_{\rm ch}$/d$\eta)^{1/3}$ for a given centrality class are taken from \cite{Aamodt:2010cz,Abelev:2008ab}.
%\cite{Gavin:2006xd} 
The STAR and ALICE collaborations estimated shear viscosities, using Eq.~\ref{eq:sigmacentral}, based exclusively on most central and most peripheral collisions. In this letter, the collision centrality dependencies of the longitudinal widths reported by both experiments are  used to investigate whether $\eta/s$ exhibits a dependence on (${\rm d}N_{\rm ch}$/d$\eta)^{1/3}$. The longitudinal broadening of the $G_2$ correlator, defined below, is expected to be insensitive to initial state density fluctuations in the transverse plane. As such, it powerfully complements studies of  $\eta/s$ based on measurements of anisotropic flow that suffer in part from such a dependence~\cite{Bernhard:2019bmu}. Measurements of $G_2$ correlators additionally have a different sensitivity to non-flow effects which make them an invaluable tool in the understanding of the dynamics of A--A collisions, and as such, provide additional testing grounds of hydrodynamical and other types of theoretical models. 

The ALICE measurements were reported in terms of a dimensionless variant of the $G_2$ correlator~\cite{Gavin:2006xd,Sharma:2008qr} defined as 
\begin{equation}
\begin{aligned}
G_2 \left(\eta_1,\varphi_1,\eta_2,\varphi_2 \right)  &= 
 \frac{1}{\langle p_{\rm T,1} \rangle \langle p_{\rm T,2} \rangle} \times\\
    & \quad \left[\frac{S(\eta_1,\varphi_1,\eta_2,\varphi_2) }{\langle n_{1} \rangle \langle n_{2} \rangle}
        - \langle p_{\rm T,1}\rangle \langle p_{\rm T,2}\rangle
    \right]
  \label{eq:gavinG}
\end{aligned}  
\end{equation}
with
\begin{equation}
S(\eta_1,\varphi_1,\eta_2,\varphi_2) =  \left\langle \sum\limits_{\text{i}}^{n_1} \sum\limits_{\text{j} \neq \text{i}}^{n_{2}} 
        p_{\text{T},\text{i}} \; p_{\text{T},\text{j}}\right\rangle
\end{equation}
where $n_1\equiv n(\eta_1,\varphi_1)$ and $n_2\equiv n(\eta_2,\varphi_2)$ are the number of charged particle tracks detected, in each event, within bins centered  at $\eta_1,\varphi_1$ and $\eta_2,\varphi_2$, respectively. Sums are carried over particle transverse momenta 
$p_{{\rm T},i}$, $i \in [1,n_{1}]$, and $p_{{\rm T},j}$, $j \neq i \in [1,n_{2}]$, respectively. The bracket notation $\langle O\rangle$ is used to represent event ensemble averages computed within the bins $\eta_i,\varphi_i$, $i=1,2$. Thus $\langle n_i\rangle $ and  
$\langle p_{\rm T,i}\rangle$ represent average number of particles and average transverse momenta in bin $\eta_i,\varphi_i$, respectively. 
The reported ALICE measurement was limited to charged particles  with transverse momenta in the range $0.2 \le p_{\rm T} < 2$  GeV/$c$ and  pseudorapidities within $|\eta|<0.8$.

\begin{figure}[ht]
\centering

\includegraphics[width=\columnwidth,keepaspectratio=true,clip=true,trim=5pt 0pt 50pt 30pt]{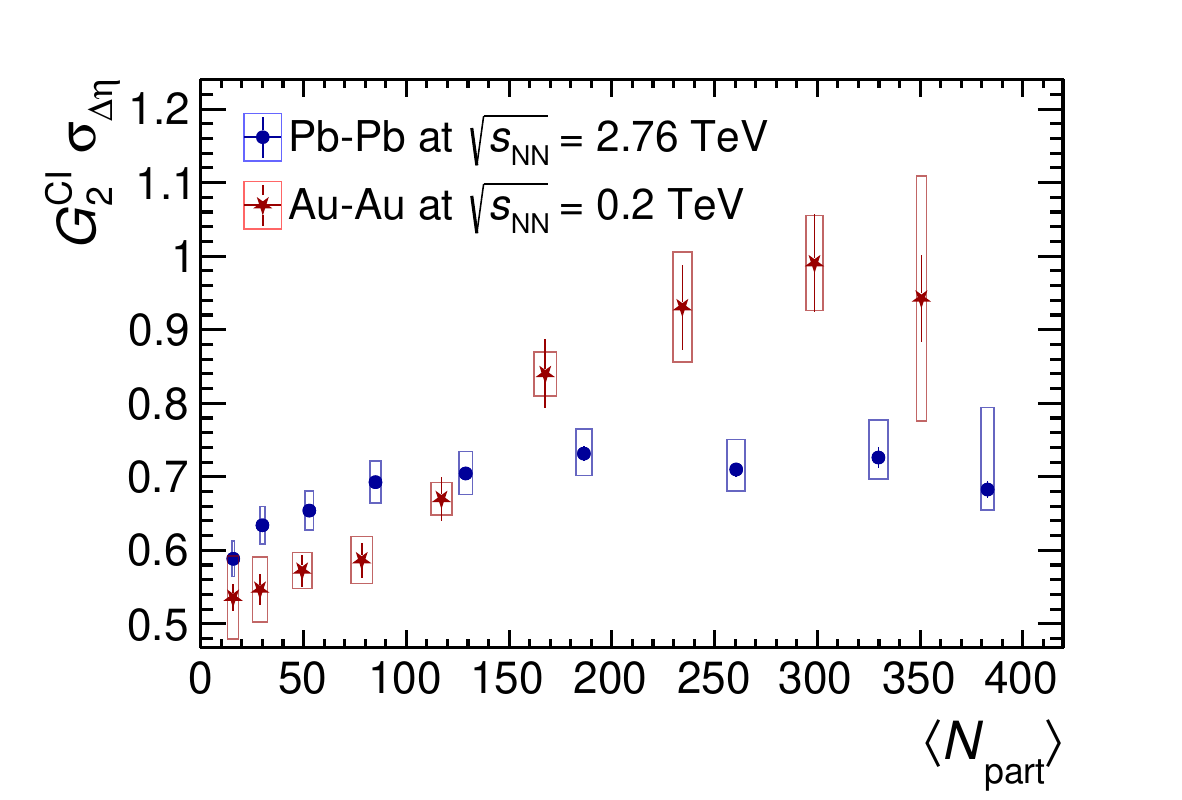}
    \caption{Longitudinal width, $\sigma_{\Delta\eta}$, of the $G_{2}^{\rm {CI}}$ correlator vs. the estimated 
  number of participants measured  in Au--Au collisions at $\sqrt{s_{\rm NN}} = 200\;\text{GeV}$~\cite{Agakishiev:2011fs}  and in Pb--Pb collisions at $\sqrt{s_{\rm NN}}=2.76\;\text{TeV}$~\cite{Acharya:2019oxz}, reported by the STAR and ALICE collaborations, respectively. Error bars and error boxes represent statistical and systematic uncertainties, respectively. %Reproduced from \cite{Acharya:2019oxz}.
  }
    \label{fig:starcomp} 
\end{figure}

 Within the context of the ALICE analysis, the $G_2^{\text CI}$ correlator dependence on $\Delta\eta$ and $\Delta\varphi$ was parametrized with a two-component model defined as
\begin{equation}
\begin{aligned}
          &F( \Delta \eta, \Delta \varphi) = B + \displaystyle\sum_{n=2}^{6} a_{n} \times \cos \left(n \Delta \varphi\right) \\& \quad +
  A \frac{\gamma_{\Delta \eta}}{2\,\omega_{\Delta \eta}\,
      \Gamma \left(\frac{1}{\gamma_{\Delta \eta}}\right)}
      \,\rm{e}^{- \left| \frac{\Delta \eta}{\omega_{\Delta \eta}}\right|^{\gamma_{\Delta \eta}}} 
          \frac{\gamma_{\Delta \varphi}}{2\,\omega_{\Delta \varphi}\,
      \Gamma \left(\frac{1}{\gamma_{\Delta \varphi}}\right)}
      \,\rm{e}^{- \left| \frac{\Delta \varphi}{\omega_{\Delta \varphi}}\right|^{\gamma_{\Delta \varphi}}}
\label{eq:fitfunc}
\end{aligned}
\end{equation}
where $B$ and $a_n$  describe the  long-range mean correlation strength and azimuthal anisotropy, respectively, while the bidimensional generalized Gaussian, whose shape is determined by the parameters $A$, $\omega_{\Delta \eta}$, $\omega_{\Delta \varphi}$, $\gamma_{\Delta \eta}$ and $\gamma_{\Delta \varphi}$, is used to model the correlation signal of interest. 
The ALICE collaboration  reported longitudinal widths $\sigma_{\Delta \eta }$  computed as the standard deviation of the generalized Gaussian 
\begin{equation}
  \sigma_{\Delta \eta } = \sqrt{\frac{\omega^2_{\Delta \eta } 
      \Gamma(3/\gamma_{\Delta \eta })}{\Gamma(1/\gamma_{\Delta \eta })}}
\end{equation}
along $\Delta\eta$~\cite{Acharya:2019oxz}. These values are plotted  as a function of the number of collision participants estimated from Glauber models~\cite{Abelev:2013qoq} in Fig.~\ref{fig:starcomp}.

Instead of using a fitting procedure, the STAR collaboration estimated the longitudinal width of measured correlators by computing the rms of one-dimensional projections of $G_2$ correlators onto the $\Delta\eta$ axis~\cite{Agakishiev:2011fs}. These rms width values are plotted vs. the number of collision participants in Fig.~\ref{fig:starcomp}.

\begin{figure}[hbt]
    \centering
\includegraphics[width=\columnwidth,keepaspectratio=true,clip=true,trim=5pt 0pt 50pt 30pt]{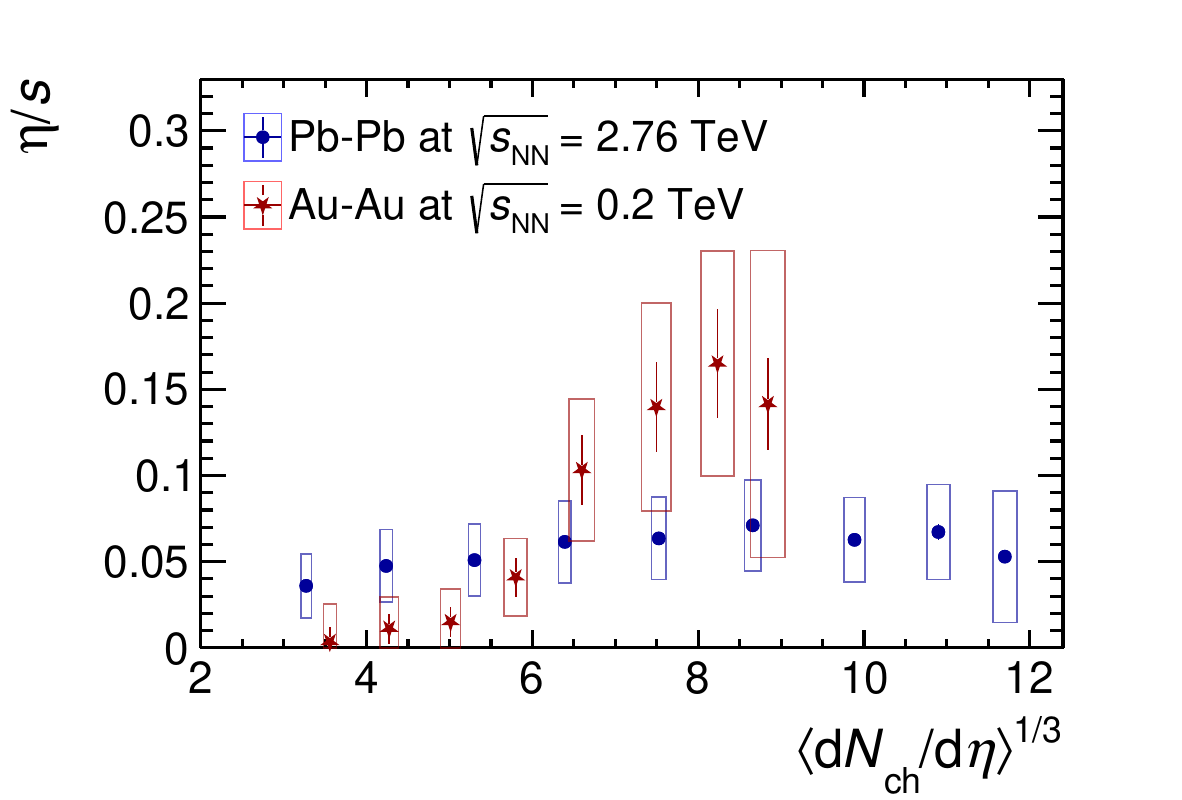}
 \caption{Values of the shear viscosity per unit of entropy density, $\eta/s$, computed in this work,  as a function of the cubic root of the charged particle density ${\rm d}N_{\rm ch}/{\rm d}\eta$ measured in Pb--Pb collisions at $\sqrt{s_{\rm NN}}=2.76\;\text{TeV}$~\cite{Acharya:2019oxz} and in Au--Au collisions at $\sqrt{s_{\rm NN}} = 200\;\text{GeV}$~\cite{Agakishiev:2011fs}. Error bars and error boxes represent statistical and systematic uncertainties, respectively. }
\label{fig:etasvsdndeta} 
\end{figure}

All parameters (d$N_{\rm ch}$/d$\eta$,  $G_{2}^{\rm CI}$ widths and $\tau_{\rm f}$) used in the computation of $\eta/s$ as well as the extracted values of  $\eta/s$ at RHIC and LHC energies are listed in Tab.~\ref{tab:paramForeta} as a function of collision centrality expressed in terms  of the fractional cross section: the range 0--5\% corresponds to most central collisions while ranges 5--10\%, 10--20\%, etc,  represent collisions with increasingly larger impact parameters. Our analysis, based on  STAR and ALICE is  limited to quasi-peripheral collisions up to the range 70--80\%, beyond which the applicability of Gavin's model might be put into question. Estimates of $\eta/s$ computed with  Eq.~\ref{eq:sigmacentral} based on the above widths and freeze-out times are listed, for both RHIC and LHC energies, in the two right-most columns of Tab.~\ref{tab:paramForeta} . The values of  $\eta/s$  vs  $({\ rm d}N_{\rm ch}/{\rm d}\eta)^{1/3}$  are plotted in Fig.~\ref{fig:etasvsdndeta}. Statistical and systematic uncertainties, the last ones incorporating the uncertainties from the fit procedure for freeze-out times extraction, are displayed with vertical bars and rectangular boxes, respectively.

\begin{table*}[htb]
    \centering
    \begin{tabular}{| c ||  c| c| c|c |} 
    \hline
   Centrality    & ${\rm d}N_{\rm ch}/{\rm d}\eta$  &       $G^{\rm CI}_2$ $\sigma_{\Delta \eta} $ &  $\tau_{\rm f}$ (fm/$c$)& $\eta/s$ \\ 
   \hline 
   \multicolumn{5}{|c|}{LHC}\\
   \hline
 0-- 5\% & 1601.00 $\pm$ 60.00 &  0.68 $\pm$ 0.01$^{\rm sta}$ $+$ 0.11$^{\rm sys}$ $-$ 0.03$^{\rm sys}$& $10.33 \pm 0.10 \pm 1.03^{\rm fit}$ & $0.05 \pm 0.00^{\rm sta} \pm 0.04^{\rm sys}$\\
 5--10\% & 1294.00 $\pm$ 49.00 &  0.73 $\pm$ 0.01$^{\rm sta}$ $+$ 0.05$^{\rm sys}$ $-$ 0.03$^{\rm sys}$& $9.62 \pm 0.09 \pm 0.91^{\rm fit}$ & $0.07 \pm 0.00^{\rm sta} \pm 0.03^{\rm sys}$\\
10--20\% & 966.00 $\pm$ 37.00 &  0.71 $\pm$ 0.01$^{\rm sta}$ $+$ 0.04$^{\rm sys}$ $-$ 0.03$^{\rm sys}$& $8.73 \pm 0.08 \pm 0.75^{\rm fit}$ & $0.06 \pm 0.00^{\rm sta} \pm 0.02^{\rm sys}$\\
20--30\% & 649.00 $\pm$ 23.00 &  0.73 $\pm$ 0.01$^{\rm sta}$ $+$ 0.03$^{\rm sys}$ $-$ 0.03$^{\rm sys}$& $7.64 \pm 0.07 \pm 0.56^{\rm fit}$ & $0.07 \pm 0.00^{\rm sta} \pm 0.03^{\rm sys}$\\
30--40\% & 426.00 $\pm$ 15.00 &  0.70 $\pm$ 0.01$^{\rm sta}$ $+$ 0.03$^{\rm sys}$ $-$ 0.03$^{\rm sys}$& $6.64 \pm 0.06 \pm 0.39^{\rm fit}$ & $0.06 \pm 0.00^{\rm sta} \pm 0.02^{\rm sys}$\\
40--50\% & 261.00 $\pm$  9.00 &  0.69 $\pm$ 0.01$^{\rm sta}$ $+$ 0.03$^{\rm sys}$ $-$ 0.03$^{\rm sys}$& $5.64 \pm 0.05 \pm 0.23^{\rm fit}$ & $0.06 \pm 0.00^{\rm sta} \pm 0.02^{\rm sys}$\\
50--60\% & 149.00 $\pm$  6.00 &  0.65 $\pm$ 0.01$^{\rm sta}$ $+$ 0.03$^{\rm sys}$ $-$ 0.03$^{\rm sys}$& $4.68 \pm 0.04 \pm 0.13^{\rm fit}$ & $0.05 \pm 0.00^{\rm sta} \pm 0.02^{\rm sys}$\\
60--70\% &  76.00 $\pm$  4.00 &  0.63 $\pm$ 0.01$^{\rm sta}$ $+$ 0.03$^{\rm sys}$ $-$ 0.03$^{\rm sys}$& $3.74 \pm 0.04 \pm 0.20^{\rm fit}$ & $0.05 \pm 0.00^{\rm sta} \pm 0.02^{\rm sys}$\\
70--80\% &  35.00 $\pm$  2.00 &  0.59 $\pm$ 0.01$^{\rm sta}$ $+$ 0.02$^{\rm sys}$ $-$ 0.02$^{\rm sys}$& $2.89 \pm 0.03 \pm 0.33^{\rm fit}$ & $0.04 \pm 0.00^{\rm sta} \pm 0.02^{\rm sys}$\\

  \hline
   \multicolumn{5}{|c|}{RHIC}\\
   \hline
   
 0-- 5\% & 691.00 $\pm$ 49.00 & 0.94 $\pm$ 0.06$^{\rm sta}$ $\pm$ 0.17$^{\rm sys}$ & $7.24 \pm 0.07 \pm 0.50^{\rm fit}$ & $0.14 \pm 0.03^{\rm sta} \pm 0.09^{\rm sys}$\\
 5--10\% & 558.00 $\pm$ 40.00 & 0.99 $\pm$ 0.07$^{\rm sta}$ $\pm$ 0.06$^{\rm sys}$ & $6.81 \pm 0.06 \pm 0.41^{\rm fit}$ & $0.16 \pm 0.03^{\rm sta} \pm 0.07^{\rm sys}$\\
10--20\% & 421.00 $\pm$ 30.00 & 0.93 $\pm$ 0.06$^{\rm sta}$ $\pm$ 0.07$^{\rm sys}$ & $6.27 \pm 0.05 \pm 0.30^{\rm fit}$ & $0.14 \pm 0.03^{\rm sta} \pm 0.06^{\rm sys}$\\
20--30\% & 287.00 $\pm$ 20.00 & 0.84 $\pm$ 0.05$^{\rm sta}$ $\pm$ 0.03$^{\rm sys}$ & $5.62 \pm 0.04 \pm 0.17^{\rm fit}$ & $0.10 \pm 0.02^{\rm sta} \pm 0.04^{\rm sys}$\\
30--40\% & 195.00 $\pm$ 14.00 & 0.67 $\pm$ 0.03$^{\rm sta}$ $\pm$ 0.02$^{\rm sys}$ & $5.05 \pm 0.04 \pm 0.12^{\rm fit}$ & $0.04 \pm 0.01^{\rm sta} \pm 0.02^{\rm sys}$\\
40--50\% & 126.00 $\pm$  9.00 & 0.59 $\pm$ 0.02$^{\rm sta}$ $\pm$ 0.03$^{\rm sys}$ & $4.48 \pm 0.05 \pm 0.17^{\rm fit}$ & $0.01 \pm 0.01^{\rm sta} \pm 0.02^{\rm sys}$\\
50--60\% &  78.00 $\pm$  6.00 & 0.57 $\pm$ 0.02$^{\rm sta}$ $\pm$ 0.02$^{\rm sys}$ & $3.95 \pm 0.06 \pm 0.27^{\rm fit}$ & $0.01 \pm 0.01^{\rm sta} \pm 0.02^{\rm sys}$\\
60--70\% &  45.00 $\pm$  3.00 & 0.55 $\pm$ 0.02$^{\rm sta}$ $\pm$ 0.04$^{\rm sys}$ & $3.43 \pm 0.08 \pm 0.38^{\rm fit}$ & $0.003 \pm 0.009^{\rm sta} \pm 0.022^{\rm sys}$\\
\hline
\end{tabular}
    \caption{Compilation of measured charged particle densities, ${\rm d}N_{\rm ch}/{\rm d}\eta$, and longitudinal widths,  $\sigma_{\Delta\eta}$, of the $G_2^{\rm CI}$ correlator, interpolated freeze-out times, $\tau_{\rm f}$, and computed values of $\eta/s$ as a function of the centrality of Pb--Pb collisions  at $\sqrt{s_{\rm NN}} = 2.76\;{\rm TeV}$~\cite{Abbas:2013bpa,Aamodt:2011mr,Acharya:2019oxz,Aamodt:2010cz} and Au--Au at $\sqrt{s_{\rm NN}} = 200\;{\rm GeV}$~\cite{Abelev:2008ab,Agakishiev:2011fs}}
    \label{tab:paramForeta}
\end{table*}

\begin{figure}[hbt]
    \centering
\includegraphics[width=\columnwidth,keepaspectratio=true,clip=true,trim=0pt 0pt 40pt 0pt]{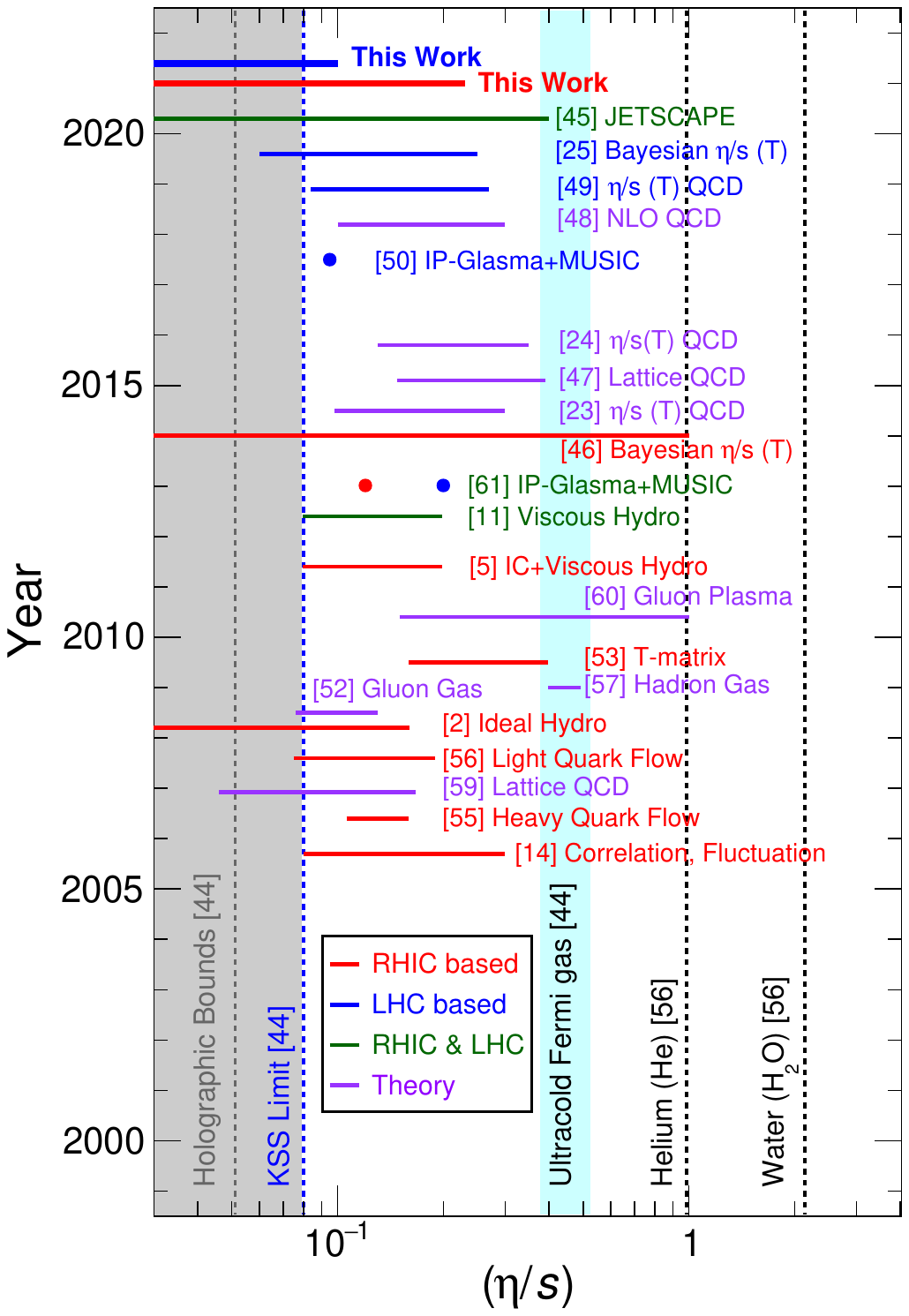}      \caption{Comparison of $\eta/s$ obtained in this work with a collection of values published since 2005 \cite{Romatschke:2007mq,Song:2010mg,Song:2012tv,Gavin:2006xd,Bernhard:2019bmu,Adams:2012th,Everett:2020xug,Novak:2013bqa,Haas_2014,Christiansen_2015,Mages:2015rea,Ghiglieri:2018dib,Dubla:2018czx,McDonald:2016vlt,DiasdeDeus:2012uc,Xu:2007ns,vanHees:2008gj,Dong:2019byy,Adare:2006nq,Lacey:2006bc,Demir:2008tr,Arnold:2003zc,Meyer:2007ic,Chen:2009sm,Gale:2012rq}, including both theoretical values and experimental values obtained from direct comparisons of theory or models results to data. Horizontal lines represent values ranges which also incorporate uncertainties. Blue and red lines correspond to values determined using LHC or RHIC data, respectively, green lines when both were used, and purple for pure theoretical results. See text for more details. }
    \label{fig:etasvsyear} 
\end{figure}

Computed values of $\eta/s$ range from $0.04 \pm 0.02^{\rm sys}$ to  $0.07 \pm 0.03^{\rm sys}$ and from a value compatible with 0  to $0.16 \pm 0.03^{\rm sta} \pm 0.07^{\rm sys}$ for LHC energies and RHIC energies, respectively. One observes that  values extracted from Pb--Pb collisions at the LHC \ \  exhibit \ \  a weak dependence \  \ on  $({\rm d}N_{\rm ch}/{\rm d}\eta)^{1/3}$, while those from Au--Au collisions, measured at RHIC, show a rising  trend with increasing $($d$N_{\rm ch}$/d$\eta)^{1/3}$, albeit with large uncertainties.  Values of $\eta/s$ obtained from Pb--Pb collisions  are close but somewhat  lower than the KSS bound  of 1/4$\pi$ , while those obtained from Au--Au collisions  are compatible with vanishing viscosities in the range $($d$N_{\rm ch}$/d$\eta)^{1/3} < 5 $ but exceed \ \  the KSS\  \   bound \ \ above $($d$N_{\rm ch}$/d$\eta)^{1/3} \approx 6$. However,  $\eta/s$ values derived from STAR and ALICE data  are compatible with one another at the one $\sigma$ level at all values of $($d$N_{\rm ch}$/d$\eta)^{1/3}$.   

\section{Discussion}
A compilation of $\eta/s$ values obtained in this work as well as  those reported in theoretical and phenomenological calculations is presented in Fig.~\ref{fig:etasvsyear}. 
Shear viscosity values for ultracold Fermi gas~\cite{Adams:2012th}, Helium ~\cite{Lacey:2006bc} and water~\cite{Lacey:2006bc}, evaluated at their respective critical temperatures, as well as the holographic bounds~\cite{Everett:2020xug} and the KSS limit~\cite{Kovtun:2004de}, have also been incorporated as baselines and references.

The results reported in this work are displayed in Fig.~\ref{fig:etasvsyear}  with two horizontal lines (one for the LHC and one for RHIC). The ranges of the horizontal lines span  the smallest to highest values obtained at each energy,  including systematic uncertainties.
Somewhat older  compilations have also been reported~\cite{Shen:2015msa}.
A comparative analysis of the results presented in this compilation  is complicated in part by the fact that viscous effects  are likely to accumulate  throughout a system's evolution. However, the shear and bulk viscosities may depend on the temperature, matter density, the presence of magnetic fields, and possibly other system conditions, that evolve as the QGP expands and goes through a transition into a hadron phase.  The model used in this work and several of the   calculations listed in the compilation neglect such a  time/tem\-pe\-ra\-tu\-re dependence and represent the viscosity as a single effective value, while others attempt to account for time and temperature dependencies using various prescriptions. The  horizontal span of the lines displayed in Fig.~\ref{fig:etasvsyear} is thus meant to represent either the range of effective $\eta/s$ values constrained by comparisons with experimental data or the ranges of $\eta/s$ values considered in  the models and yielding a good representation of the measured data. Although in both cases the reported uncertainties have been also incorporated into to the line length, the apparent relationship should not be considered as a statement of the precision achieved in the studies included in this compilation.

Comparative studies of hydrodynamics and measured data arguably culminated with studies based on a Bayesian estimation of the properties of the QGP~\cite{Bernhard:2019bmu,Adams:2012th,Everett:2020xug,Novak:2013bqa} yielding most probable  $\eta/s$ values in rather good agreement with the results of this work.
Estimates of $\eta/s$ obtained in  this work are also in quantitative agreement with  QCD inspired calculations including, for instance, estimates based on nonperturbative gluon spectral functions at finite temperature in quenched QCD with the maximum entropy method~\cite{Haas_2014} and calculations based on the Kubo formula in Yang-Mills theory~\cite{Christiansen_2015}. 
Results of this work are also in qualitative agreement with estimates based on lattice QCD (LQCD) calculations of the QGP transport coefficients~\cite{Mages:2015rea} and perturbative QCD calculations at almost NLO \cite{Ghiglieri:2018dib}, as well as recent calculation based on the MUSIC framework that used a temperature dependent $\eta/s$ computed with a QCD based approach \cite{Dubla:2018czx}. In all cases in which the temperature evolution of $\eta/s$ have been estimated the range reported in Fig.~\ref{fig:etasvsyear} goes up to $T=3T_{\rm c}$ which in most of the cases matches the published range.   
It is interesting to address the results reported in this work from an additional perspective. Although as was mentioned before the Gavin ansatz does not consider any temperature dependence for $\eta/s$ the temperature reached by the medium produced in A--A collisions, presumably, will not be the same for the different centrality ranges. The values of $\eta/s$ quoted in this work for each centrality range at both energies (i.e., RHIC and the LHC) thus correspond  to an effective viscosity which condensates the whole system evolution for that centrality range. If as suggested above, the produced  system reach different temperatures in each centrality range, there could be an implicit link between the results presented in this work and the evolution of $\eta/s$ with system temperature.
 
Taking the estimates of $\eta/s$  shown in Fig.~\ref{fig:etasvsdndeta} at face value, it is interesting to consider whether they might have any implications concerning the nature and properties of hot QCD matter produced in heavy-ion collisions at RHIC and LHC. First consider that estimates of the  initial temperature reached in central Pb--Pb collisions at LHC suggest it is of the order of 300 MeV, i.e., 30\% larger compared to that achieved at RHIC in central Au--Au collisions~\cite{Busza:2018rrf,Karsch:2001cy,Muller:2012zq,Martinez:2013xka}. 
Also consider that
the fireball formed in Pb--Pb  collisions at the LHC have been estimated to  live approximately 40\% longer that those produced in Au--Au collisions at RHIC~\cite{Aamodt:2011mr}. This implies that shear viscous forces have more time to operate in central  Pb--Pb collisions at LHC than in Au--Au at RHIC. For systems of equal $\eta/s$ and  temperature, one would
expect to observe a larger longitudinal broadening of the $G_2$ correlator in Pb--Pb but the observed broadening is in fact smaller than that seen in central Au--Au collisions. Taken at face value, this suggests that the effective shear viscosity per unit of entropy is smaller in Pb--Pb at 2.76 TeV. We should stress, however, that the extracted values of $\eta/s$ reflect  the complete evolution of  systems formed in A--A collisions. It is consequently incorrect to associate  and use a particular system temperature to evaluate the shear viscosity. Indeed, estimates should account for possible  evolution of $\eta/s$ with temperature explicitly or be based on an appropriate effective, time averaged, system temperature. The interpretation of the data is further complicated by the likely presence
of kinematic narrowing associated to radial flow. The average transverse momentum, $\langle p_{\rm T}\rangle$, is found to be approximately 10\% larger at LHC energies compared to RHIC. This increase may in part result from faster radial flow at the TeV energy scale. It is well established that strong radial flow produces a sizable narrowing of two particle correlators, such as balance functions $B$~\cite{Adams:2003kg,Aggarwal:2010ya,Abelev:2013csa,Adam:2015gda}, as well as generic number and transverse momentum correlators $R_2$ and $P_2$~\cite{Acharya:2018ddg}, respectively. A similar narrowing is thus expected also for $G_2$ and has in fact been found to occur in Pb--Pb collisions: the $G_2^{\rm CD}$ correlator, in particular, exhibits a significant narrowing from peripheral to central Pb--Pb collisions reported by the ALICE collaboration~\cite{Acharya:2019oxz}. While this narrowing is most easily and explicitly observed for unlike-charge particle pairs, it should also be occurring for like-sign  pairs contributing to the $G_2^{\rm CI}$ correlator. Narrowing effects associated with kinematic focusing might then partially counterbalance the broadening due to viscous forces, and thus effectively reduce values of $\eta/s$ extracted from both the ALICE and STAR data. But given the radial flow is likely somewhat stronger at LHC, that could imply the difference seen between central Au--Au and Pb--Pb collisions is in part due to the presence of extra focusing at LHC energy. 

Additional theoretical calculations~\cite{Karpenko:2015xea}  suggest that  $\eta/s$ should increase with decreasing collision energy within the RHIC energy domain in part as a result of an explicit dependence on the matter baryochemical potential $\mu_{\rm B}$~\cite{Auvinen:2017fjw}. Studies of relative yields of produced hadrons indicate that the baryochemical potential is nearly vanishing at central rapidities in Pb--Pb collisions, with  values of order $\mu_{\rm B} \sim $0.7 reported by global thermal fits  \cite{Andronic:2016nof,Andronic:2017pug}, while significantly larger values, $\mu_{\rm B} \sim $20, were  extracted based on Au--Au collisions at RHIC top energy~\cite{Andronic:2005yp}. Differences of $\eta/s$ observed by STAR and ALICE collaborations might thus also result in part from this  change of the baryochemical potential. While we note that  the precision of the data is clearly insufficient to establish any firm conclusion on such a dependence, we stress that precise  measurements of the $G_2$ correlator in the context of the second RHIC beam energy scan (BES-II) might in fact provide better grounds to seek evidence of this dependence.  Studies of the $G_2$ correlator with RHIC beam energy scan data are thus indeed of high interest. 

Other considerations are also of interest. Collisions of large nuclei at ultra-high energy, both at RHIC and LHC, are expected to produce very large magnetic fields and have been predicted to induce large vorticity and global polarization effects as well as finite out of plane charge separation associated with the chiral magnetic effect (CME). While the existence of the CME remains to be established, both STAR and ALICE collaborations have reported observations of global polarization of $\Lambda$-baryons \cite{Acharya:2019ryw,STAR:2017ckg,Adam:2018ivw} believed to result from the presence of large vorticity in Au--Au and Pb--Pb collisions. It has been suggested that the presence of  large magnetic fields might also have an impact on viscous effects~\cite{MCINNES2018455} as they might strongly suppress momentum diffusion in the reaction plane or impart a ``paramagnetic squeezing'' effect capable of  altering pressure gradients. Variations of the magnetic field strength and its time evolution as function of centrality and collision energy,  may also influence the effective diffusivity and the viscosity of the QCD matter produced in these collisions. The magnitude of the effect is as of yet unknown but nonetheless worthy of  additional investigations given  the current uncertainties in  $\eta/s$ values do not allow to conclude about a possible difference at the two energies.

Additionally, in order to make progress on a full characterization of $\eta/s$ as a function of temperature, collision energy, baryochemical potential, etc., additional and more precise measurements of  $G_2^{\rm CI}$  at different collision energies are necessary; for example, as already mentioned, from the BES-II at RHIC. Furthermore, supporting theoretical studies in the framework of relativistic hydrodynamics will also be greatly beneficial. One needs, in particular, to establish the influence of the temperature and viscosity of the different stages of the collision (QGP, phase transition region, hadronic phase) on the longitudinal broadening of  $G_2^{\rm CI}$. The role of resonance decays and  charge conservation must also be clarified in association with quantitative studies of the radial flow velocities imparted to the matter produced in A--A collisions. Ideally, these studies should be conducted for several system sizes and collision energies.  In light of observations of $\Lambda$-baryon global polarization already mentioned, it shall also be of interest to examine whether the strong  magnetic fields present at the onset of A--A collisions can persist long enough to have a quantitatively measurable impact on the shear viscosity in general, and on the longitudinal broadening of transverse momentum correlators in particular. 

\section{Conclusion}
We presented an evaluation of the collision centrality dependence of the shear viscosity per unit of entropy, $\eta/s$ of the Quark Gluon Plasma produced in A--A collisions at RHIC and LHC based on measurements of the $G_2$ correlator by the STAR and ALICE collaborations using the Gavin ansatz embodied in Eq.~\ref{eq:sigmacentral}. Freeze-out times required to carry out the calculations were determined  as a function of the cubic root of the charged particle multiplicity (or collision centrality) from two-pion Bose-Einstein measurements. Values of $\eta/s$ obtained in Pb--Pb collisions, based on ALICE data, indicate the shear viscosity per unit of entropy is of the order of the KSS bound and essentially independent of collision centrality at LHC energy. By contrast, the STAR data are consistent with vanishing $\eta/s$ values in peripheral collisions and values exceeding the KSS bound in more central collisions. However, given the large systematic uncertainties of these data, one cannot exclude $\eta/s$ might be invariant with collision centrality. Likewise, one cannot readily exclude that values of $\eta/s$ might  also be invariant with beam energy.

The precision of our estimates of the dependence of $\eta/s$, particularly at RHIC, are limited by the accuracy of the STAR measurement of $G_2$. Uncertainties are also largely determined by the various caveats associated with the Gavin ansatz discussed above, most particularly the choice of characteristic temperature used in the calculation. Clearly, a more detailed calculation along the lines of Ref.~\cite{Gavin:2016jfw} are needed to improve on this work.

Finally, we stress that although measurements of $G_2$ are challenging, owing in particular to their sensitivity to $p_{\rm T}$ dependent efficiency corrections, they are nonetheless possible as demonstrated by the recent ALICE measurement. Precise studies of the evolution of  the $G_2^{\rm CI}$ correlator with collision centrality thus stand to become a discriminating gauge of not only the average magnitude of the shear viscosity per unit of entropy, $\eta/s$, but its temperature dependence also. As such, they might provide new and valuable inputs to multi-system Bayesian constraints methods. Ideally, this will require measurements of $G_2$ be completed based, for instance, on the beam energy scan at RHIC as well as for  smaller collision systems at the LHC and RHIC.
%%%%% acknowledgements

\section*{Acknowledgements}
SB acknowledge the support of the Swedish Research Council (VR). This work  was also supported in part by the United States Department of Energy, Office of Nuclear Physics (DOE NP), United States of America, under grant No.  DE-FG02-92ER40713.

\bibliographystyle{spphys}
\interlinepenalty=10000
\bibliography{bibliography}
\end{document}